\documentclass[sigconf]{acmart}

\AtBeginDocument{%
  }

\copyrightyear{2025}
\acmYear{2025}
\acmConference[CHI '25]{CHI Conference on Human Factors in Computing
Systems}{April 26-May 1, 2025}{Yokohama, Japan}
\acmBooktitle{CHI Conference on Human Factors in Computing Systems (CHI'25), April 26-May 1, 2025, Yokohama, Japan}
\acmDOI{10.1145/3706598.3713150}
\acmISBN{979-8-4007-1394-1/25/04}

\usepackage{arabtex}
\usepackage{utf8}
\usepackage{tabularx}

\usepackage{subcaption}

\usepackage{multirow}
\usepackage{multicol}
\usepackage{enumitem}

\begin{document}

\title[Who Should Set the Standards?]{Who Should Set the Standards? Analysing Censored Arabic Content on Facebook during the Palestine-Israel Conflict}

\author{Walid Magdy}
\email{wmagdy@inf.ed.ac.uk}
\orcid{0000-0001-9676-1338}
\affiliation{
  \institution{The University of Edinburgh}
  \city{Edinburgh}
  \country{UK}
}

\author{Hamdy Mubarak}
\email{hmubarak@hbku.edu.qa}
\orcid{0000-0003-4828-6098}
\affiliation{
  \institution{Qatar Computing Research Institute}
  \city{Doha}
  \country{Qatar}
}
\author{Joni Salminen}
\email{jonisalm@uwasa.fi}
\orcid{0000-0003-3230-0561}
\affiliation{
  \institution{University of Vaasa}
  \city{Vaasa}
  \country{Finland}
}
%

\renewcommand{\shortauthors}{Magdy et al.}

\begin{abstract}

Nascent research on human-computer interaction concerns itself with fairness of content moderation systems.
Designing globally applicable content moderation systems requires considering historical, cultural, and socio-technical factors.
Inspired by this line of work, we investigate Arab users' perception of Facebook's moderation practices. We collect a set of 448 deleted Arabic posts, and we ask Arab annotators to evaluate these posts based on (a) Facebook Community Standards (FBCS) and (b) their personal opinion. Each post was judged by 10 annotators to account for subjectivity. Our analysis shows a clear gap between the Arabs' understanding of the FBCS and how Facebook implements these standards. The study highlights a need for discussion on the moderation guidelines on social media platforms about who decides the moderation guidelines, how these guidelines are interpreted, and how well they represent the views of marginalised user communities.

\end{abstract}



\maketitle

\section{Introduction}

Social media platforms, such as Facebook (FB), X, and YouTube, are increasingly associated with politics \cite{margetts_political_2017,koiranen_changing_2020}. People discuss politics, donate money and effort to political causes, share and consume political news, engage with political parties, and signal their political beliefs on social media platforms \cite{margetts_political_2017,aldayel2019your,kumpel_matthew_2020}. As such, these platforms exert social influence by aggregating news \cite{lerman_social_2007}, deciding what information is shown at a given time, and which user voices are amplified and which are silenced \cite{elmimouni2024shielding}. With its nearly three billion monthly active users \cite{statista_facebook_nodate}, FB is considered the largest social media platform for all users worldwide. This global popularity results in quasi-monopolistic power held by FB which, if abused, could stifle the diversity of political opinion and users' digital rights, deterring online participation and freedom of expression \cite{awwad2024digital}. 

In general, users' perception of the bias and fairness of the social media platforms matters for the design of inclusive platforms where all participants, regardless of their race, gender, culture, language, or political beliefs, feel welcome.
To this end, in-depth understanding of user perceptions of inclusion on these global platforms produces useful knowledge for betterment of online experiences, which matters for the human-computer interaction (HCI) community, in which inclusivity is considered a core value.

In terms of terminology, \textit{bias} in social media refers to systematic skewing of user participation and content toward certain demographics, viewpoints, or interests due to platform design, algorithmic amplification, and self-selection of users \cite{jiang_reasoning_2020,johnson_algorithmic_2020}. \textit{Fairness} in social media moderation specifically refers to the equal and transparent treatment of different users, points of view, and content in terms of visibility, reach, and platform privileges, regardless of demographics, political stance, or social status \cite{binns_fairness_2018,jhaver_did_2019}. \textit{Inclusivity} in this context refers to ensuring that the platform is accessible, welcoming, and usable for people of all backgrounds, abilities, languages, and demographic groups, with features and policies that enable meaningful participation by all \cite{haimson_disproportionate_2021}. These concepts are particularly relevant in the Palestinian context, where historical power imbalances and complex geopolitical factors influence how content moderation decisions affect different user groups.

One of the main occasions where substantial discussion was raised about the possibility of bias in FB was when the conflict occurred between Palestine and Israel in May 2021 (which continues to take place at the time of writing). In 2021, several news media began reporting claims about the application of systematic censorship by Facebook and the prevention of free expression~\cite{FB_censor_palestine}. 
This \textit{Palestine-Israel conflict} is summarised by the Washington Post \cite{dwoskin_facebooks_2021} as follows:

\begin{quote}
\textit{
    ``Palestinian activists took to the social media platforms as they began staging protests in late April ahead of an impending Israeli Supreme Court case over whether settlers had the right to evict families from their homes in the Jerusalem neighborhood of Sheikh Jarrah. Potential evictees live-streamed confrontations and documented footage of injuries after Israeli police stormed al-Aqsa Mosque, one of the holiest sites in Islam. The conflict descended into war after terrorist group Hamas, which governs Gaza, fired explosive rockets into Israel. Israel responded with an eleven-day bombing campaign that killed 254 Palestinians, including 66 children. Twelve people in Israel were killed, including two children. During the barrage, Palestinians posted photos on Twitter showing homes covered in rubble and children’s coffins. A cease-fire took effect May 20 [2021].''}
\end{quote}

The Palestine-Israel conflict on social media represents a unique case study in content moderation for several reasons. First, it involves deeply rooted historical narratives that shape how different groups interpret and share content. Second, the conflict highlights the challenges of moderating content in languages and cultural contexts that may be unfamiliar to platform moderators primarily based in Western countries. Third, the power dynamics between Palestinians and Israelis offline is reflected in their digital experiences, where Palestinians often face additional barriers to having their voices heard \cite{aal2024influence}. This context is crucial to understanding why content moderation decisions during the conflict became a flash point for broader discussions about fairness and bias in social media governance.
This conflict illustrates how social networks are being used in real time to communicate information and beliefs during a crisis or conflict.
In this particular conflict, there were many posts and discussions on social media, including on FB, of opposing views that supported one of the sides. Within the Arabic region, there were claims that FB was deleting Arabic posts--which were mostly supporting the Palestinian side--and restricting Arab accounts' access to the platform. The reason given by FB was that this content contradicted the platform's community standards\footnote{\url{https://transparency.fb.com/ar-ar/policies/community-standards/}}.

\begin{figure*}
\centering
\begin{minipage}[t]{.3\textwidth}
\centering
  \includegraphics[width=\linewidth]{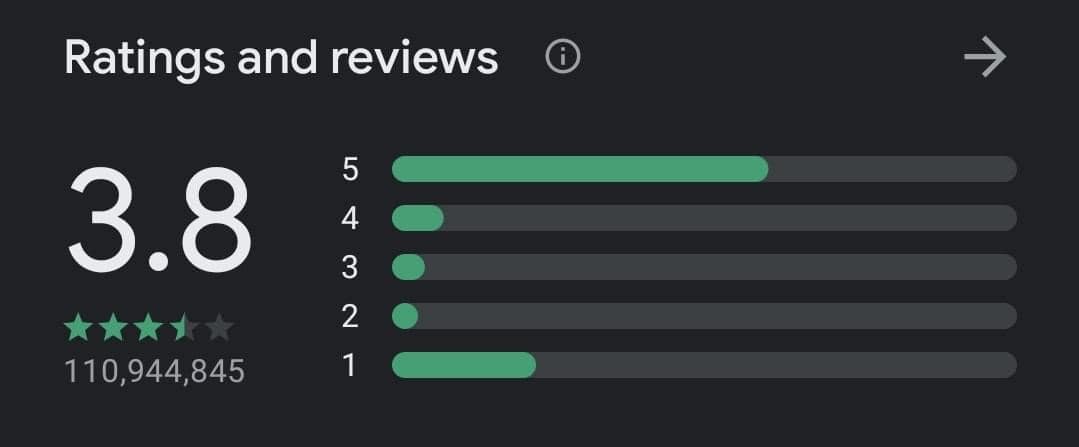}
  (a) 17 May 2021, few days after campaign start
\end{minipage}%
\hfill
\begin{minipage}[t]{.3\textwidth}
\centering
  \includegraphics[width=\linewidth]{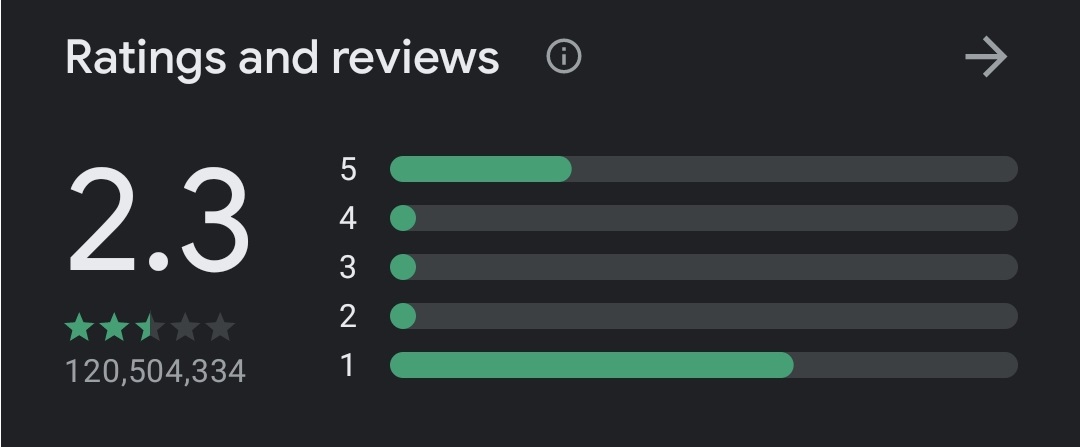}
  (b) 09 Sep 2021, four months after campaign
\end{minipage}
\hfill
\begin{minipage}[t]{.3\textwidth}
\centering
  \includegraphics[width=\linewidth]{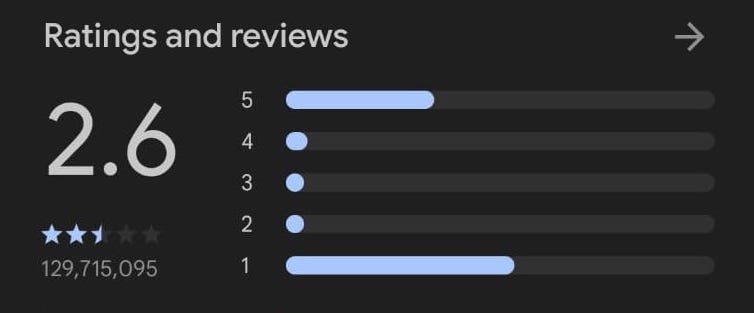}
  (c) 16 Aug 2023, 27 months after campaign
\end{minipage}
\caption{Rating of the Facebook app in Google's Play Store (screenshots). Previously, the rating was 4.0 (out of 5), but the activist campaign decreased the rating in May 2021. Screenshot (a) shows the rating in 17 May 2021, few days after the start of the campaign.  Screenshot (b) and (c) is from September 9, 2021 and August 16, 2023 respectively, implying that the damage to reputation from activist campaigns can be long-lasting.}
\label{fig:FBrating}
\end{figure*}

One of the initial reports on the topic was by the Arab Center for Social Media Advancement, recording some 500 violations of Palestinians' digital rights during this conflict event in May 2021 \cite{sneineh_facebook_2021}, which gave further rise to the \textit{perception} of bias. This perception led to a campaign by pro-Palestinians activists, particularly within the Arabic region, to give one-star ratings to FB apps in the Google Play and Apple App stores~\cite{FB_rating_1,FB_rating_2}, which, in turn, decreased FB's app rating in the Google Play Store from 4.0 to 2.3 in less than a month (see Figure \ref{fig:FBrating}). Similarly, FB's rating was significantly decreased in many regional Apple App stores, going down to less than 2.0 in many regions~\cite{FB_rating_1,FB_rating_2}. Also, more than 170 FB employees submitted a request to the company to address perceived bias~\cite{FB_employees_2021}\footnote{``As highlighted by employees, the press and members of Congress, and as reflected in our declining app store rating, our users and community at large feel that we are falling short on our promise to protect open expression around the situation in Palestine.'' \cite{staff_facebook_2021}}. The uproar about possible censoring of pro-Palestinian content motivated some Arab users to bypass FB's detection algorithms by using the ancient form of Arabic letters, where no dots are used with the letters, making the posts unreadable by machines~\cite{arabic_no_dots}. 

Although this is not the first time that FB is accused of censoring political views ~\cite{hooker2019censorship,nurik2019men,guo2020third,jackson2014censorship} (``overzealous software algorithms that are designed to protect but end up wrongly penalising marginalised groups that rely on social media to build support.'' \cite{dwoskin_facebooks_2021}), there is a lack of empirical analysis of these accusations, especially from the point of view of the user group that perceives they are being mistreated. This social media activist campaign against FB indicates a need for investigating the appropriateness of FB's content moderation practices, and whether the user community, especially one that is experiencing marginalisation, agrees with these practices. 

The controversy led Meta (the company that owns FB) to publish a report by an independent company ``Due Diligence Insights'' that analysed Facebook's impact on human rights in Israel and Palestine during the escalation in May 2021. The report listed 21 recommendations, and FB has committed to implementing 10, partially implementing four, and assessing the feasibility of the rest. The company mentioned that implementing these recommendations will take time. 
The results of this report motivate 
more research on this topic. 
To this end, Human rights watch (HRW) conducted an in-depth investigation on the topic, which led to two reports published in December 2023 \cite{hrw_meta_censor,hrw_report_meta} explicitly accusing Meta of censoring pro-Palestinian content during the conflict in 2021 and in its aftermath. In July 2024, FB released a report stating that they had banned posts containing the Arabic word ``Shahid'', which translates to ``Martyr'', and that they have lifted this ban except posts that still violate their community standards \cite{FB_transparency}.

Previous studies on social media perceptions typically focus on users from the so-called WEIRD countries (Western, Educated, Industrialised, Rich and Democratic) \cite{kumpel_matthew_2020,batool_expanding_2024}, which is a problem for understanding cross-cultural implications for the design of accessible and inclusive social media platforms from a global perspective. Because it is important to understand users beyond the WEIRD constellation, this event poses a great opportunity to study Arab users' perceptions of FB's moderation practices.

The main objective of this study is to investigate the content moderation of FB as a global social media platform that aims to be inclusive of users worldwide with diverse viewpoints.  
We put forth critical motivational questions for the HCI community:``\emph{Who should set the guidelines for content moderation on global platforms to insure diversity and inclusiveness? How should these guidelines be interpreted and implemented?}''.
In particular, we investigate 
how Arab users perceive the fairness of FB moderation during this period. 
We achieve this objective by collecting a set of deleted Arabic posts by FB, then addressing the following research questions (RQs):

\begin{itemize}
    \item \textbf{RQ1:} \textit{Do deleted Arabic posts by FB violate the platform's community standards?} This question helps to address whether FB's moderation practices work as intended.
    \item \textbf{RQ2:} \textit{How do Arab users agree on the content moderation of FB from their point of view?} This question helps to address how community standards are understood among specific groups.
    \item \textbf{RQ3:} \textit{How does the topic of moderated content affect users' (dis)agreement with the moderation?} This question helps address the possible reasons for disagreement between FB's decisions and users.    
\end{itemize}

To investigate these RQs, we developed a survey asking Arab users to provide posts that FB has deleted. We collected a set of 448 posts covering multiple topics, with the majority of them discussing the Palestine-Israel conflict. We then prepared an annotation task and asked another set of Arab users to annotate whether the posts violated the \textit{Facebook community standards} (FBCS), and to express their opinion on whether these posts should be removed or not. Our results show that the majority of annotators ($\approx$60\%) believe that these posts do not violate any of the FBCS; further, 71\% believe that these posts do not require moderation in their personal opinion. We further analyse the topics discussed in these posts and show that there is variation in understanding the FBCS among different cultures, particularly the Arabs. These results highlight the need for a discussion in the HCI community about how ``community standards'' of a global platform used by half of the population on earth should be set, understood, and applied across multiple cultures and conflicting opinions.

The current research is closely aligned with the emphasis of the HCI community on understanding people and their interactional contexts \cite{awwad2024digital,elmimouni2024shielding}. We investigate the perceptions of Arab users, offering a comprehensive exploration of their interactions with a major technological platform, FB. 
Our analysis of the unique cultural, geographic, and social context of the Arab community resonates with the interest of the HCI community in locally relevant contexts. Furthermore, collaboration between multiple annotators to judge posts emphasises the collaborative behaviours that the HCI community values \cite{souza_taxonomy_2023}.
Our findings are both timely and original, addressing the pivotal issue of digital rights violations and content moderation. By investigating how FB's moderation practices are perceived by Arab users, the study offers a fresh perspective, enriching the HCI community's understanding of how different cultural groups perceive and interact with global technological platforms. These insights have practical implications for content moderation guidelines on social media platforms, making the study a valuable contribution to both practical and theoretical domains.

The remainder of this work is organised as follows. Section 2 provides a comprehensive review of the literature on the governance of social media platforms, content moderation, and the effect of platform policies on the general user experience within the predominant platforms. Section 3 explains the methodology. Section 4 presents the data annotation process in the context of the FBCS. The results are presented in Section 5. Because it is vital to suggest concrete
recommendations for technology platforms \cite{awwad2024digital}, we do so in Section 6.
Section 7 concludes the work.

\section{Related Work}

\subsection{Moderation of Pro-Palestinian Content in Social Media}

Previous research on social media moderation during the Palestine-Israel conflict has highlighted several key themes, such as the role of platform policies in shaping narrative visibility; the impact of automated content moderation on marginalized voices; the intersection of offline power dynamics with online content restrictions; and the challenges of applying Western-centric moderation standards to Middle Eastern contexts \cite{abokhodair2024opaque, elmimouni2024shielding, aal2024influence, awwad2024digital}.

Studies have particularly focused on FB's approach to moderating Palestinian content.
\citet{abokhodair2024opaque} surveyed 200 and interviewed 12 FB users to understand their perception of FB censorship towards pro-Palestine content after the 2021 conflict, and reported similar findings to HRW reports \cite{hrw_meta_censor,hrw_report_meta} in that Arab users perceive FB as biased against the content supporting Palestine.
A follow-up study \citet{elmimouni2024shielding} showed a substantial disparity between how activists perceive the reasons for moderation and the official explanations provided. Their findings have critical implications on the potentialities of democratic discourse on social media platforms.

\citet{awwad2024digital} studied digital depression in Palestine by interviewing 19 Palestinian social media users. They found that Hamas and the Palestinian Authority engage in digital surveillance and control of Palestinian social media activists, based mainly on manual surveillance and personalised intimidation tactics. This highlights the fact that marginalised users do not only face restrictions for their freedom of expression from online platforms but also from other entities.
The researchers \cite{awwad2024digital} further identified that the Palestinian activists in their sample were often confused about why their posts were removed or accounts suspended, suggesting a lack of transparency in the decision-making process of social media platforms. Any misunderstandings of the alleged violations can easily lead to lose-lose situations in which activists become \textit{more} entrenched in their general beliefs that social networks discriminate their particular views. As a remedy, Awwad and Toyama \cite{awwad2024digital} recommend direct engagement with users, though such an approach is difficult to implement at scale due to the lack of human moderators \cite{vaccaro_at_2020,vaccaro_contestability_2021}.

Overall, although these studies point to consistent findings, they rely on qualitative analyses without analysing what exact content is censored and how this aligns (or does not) with the community standards of FB. In the current research, we try to address this gap by applying a quantitative analysis to the posts that have been removed by FB during that period.
In the following subsections, we contextualise the current research and provide the reader with a historical overview on the development of content moderation on social media. This overview can help better understand the intricacies involved with online moderation and why its implementation is not an easy task.

\subsection{The Historical Evolution of Platforms' Role in Governing Content}

There is an increasing trend of politicisation of social media \cite{koiranen_changing_2020,neubaum_its_2021,aal2024influence}. Politicisation is defined by the Oxford dictionary as ``the action of causing an activity or event to become political in character,'' or ``the process of becoming or being made politically aware.'' Both definitions manifest themselves on social media. On the one hand, political activities take place--for example, social causes that are advocated by some users \cite{waeterloos_designing_2021} and attacked by others \cite{anderson_social_2017}, resulting in online conflicts and firestorms \cite{einwiller_journalists_2017}. Users' political activities contribute to the shifting role of social media from being neutral information conveyor to being perceived as active modifier of public opinion. The logical consequence is the need for scrutiny regarding how much and by which principles social media platforms remove content, and whether their moderation policies and practices are, first of all, not violating users' digital rights \cite{mills_law_2015} and, secondly, fair in the sense that all groups and individual users undergo the same process and criteria for moderation \cite{jiang_reasoning_2020}, with a possibility of appealing about decisions that the users perceive unfair. 

The considerations of fairness are further accentuated by algorithm-based machine learning (ML) models that may inadvertently introduce bias without the knowledge of the platform's developers and designers \cite{lee_detecting_2018}. One essential question is how to govern and audit the platform's content moderation, including (1) automated process, (2) manual processes, (3) criteria fairness, and (4) the possibility of appealing about decisions that the user perceives unfair. The last point is compatible with the second part of Oxford's definition for politicisation, in that social media users are becoming more aware of the possibility of discrimination within online communities. While perceptions of bias and unfairness appear to be a nascent trend \cite{FB_rating_1,FB_rating_2}, there is a lack of studies that investigate user assessments on the appropriateness of moderation principles and how these principles are implemented by social media platforms. %

The idea of platform neutrality comes from the principle that ``all Internet traffic should be treated equally'' \cite{wikipedia_net_2021}. This concept characterised the early development of online communities \cite{croeser_post-industrial_2019}, followed by a period of commodification and commercialisation of the Internet, where terms such as user-generated content, Web 2.0, and blogosphere encapsulated a sense of positive idealism and participation \cite{glassman_logic_2011}. The Internet went mainstream and anyone was able to share their opinions freely. However, this diversity of opinions \cite{koiranen_changing_2020,neubaum_its_2021} became a hotbed for disagreement and conflict \cite{zeitzoff_how_2017,van_niekerk_social_2013}, as people of different cultural, political and religious backgrounds joined the discussion on social media platforms, sometimes with dramatically different viewpoints~\cite{aldayel2019your,HaririMagdyWolters2021}, resulting in clashes manifested by online hate \cite{salminen_anatomy_2018}, toxicity \cite{wulczyn_ex_2017}, trolling \cite{kou_flag_2021}, and cyberbullying \cite{slonje_nature_2013}. These phenomena have influenced platforms to back off neutrality and increase their moderation efforts, which in turn risks introducing new sources of bias and discrimination.

With the increase in negative side effects, online users have become acquainted with the ``dark side'' of social media \cite{salo_dark_2018}. This shift in sentiment was also reflected in research topics, with the focus shifting from the sense of optimism to the adverse effects of uncontrolled free speech, emphasising the need for detection and moderation. It became understood that the Internet could be used for negative and harmful purposes, including the propagation of extremist content \cite{ayres_cyberterrorism_2016}, radicalisation of the youth \cite{thompson_radicalization_2011}, coordinated attacks (``raids'') against other communities \cite{gerbaudo_social_2018}, spread of misinformation \cite{valenzuela_paradox_2019}, political bots \cite{woolley_social_2016}, and trolling \cite{hannan_trolling_2018,kou_flag_2021,schulenberg_towards_2023} aimed at creating chaos and confusion among users. According to the theory of network effects \cite{dou_engineering_2013}, the Internet did not create these adverse effects (``people have always associated with like-minded others'' \cite{margetts_political_2017}) but it did amplify them. Because creating connections between users is easier, when this process is influenced by similarity of opinions, homophily, that is, ``the tendency of individuals to associate and bond with similar others'' \cite{bessi2016homophily}--is likely to occur among social media users \cite{bakshy_exposure_2015,aldayel2019your,darwish2017improved,aldayel2021stance}.

The perception of social media from a positively viewed into a negatively viewed environment can be observed from critical events. For example, during the Arab spring--a social movement taking place in the Arab world in the early 2010s--social media platforms were perceived as liberators and enablers of freedom of expression \cite{khondker_role_2011}. Approximately ten years later, the escalation of the Israel-Palestine crisis in the spring 2021 mobilised a visible opposition against social media platforms' role in limiting  free expression \cite{el-gundy_ancient_2021,sneineh_facebook_2021,awwad2024digital}. 

\subsection{Content Moderation and Users' Perception}

Algorithmic content moderation aims to improve the health of the online community by filtering out offensive comments and personal attacks \cite{binns_like_2017}. Although such systems are productive \cite{lampe_slashdot_2004}, they also pose issues. First, not much is known about users' understanding and acceptance of algorithmic moderation decisions in social media \cite{kou_mediating_2020,schulenberg_towards_2023}. These understandings are possibly mediated by shared norms, values, practices, and knowledge of a community \cite{kou_mediating_2020}. Users also tend to generate folk theories about how social media moderation systems operate \cite{myers_west_censored_2018,solano-kamaiko_explorable_2024}, with possibly false or exaggerated rationale. Negative perceptions about content moderation can be related to the general idea of automation or the inconsistent behaviour of the moderation algorithm \cite{vaccaro_at_2020}.

Users do not always receive adequate explanations when the platform removes posts \cite{jhaver_does_2019}, and this lack of transparency affects user perceptions of the platform’s moderation process \cite{jhaver_did_2019}, according to the Fairness, Accountability, Trustworthiness and Control (FACT) studies on the moderation of algorithmic content \cite{vaccaro_at_2020,kou_mediating_2020,schulenberg_towards_2023,vaccaro_contestability_2021}. Furthermore, how AI moderates social media platforms is difficult for average users to understand \cite{kou_mediating_2020,schulenberg_towards_2023}. In a survey among 907 Reddit users \cite{jhaver_did_2019} whose posts had been removed, it was found that less than a fifth (18\%) considered the removal of their post(s) appropriate, while more than a third (37\%) expressed a lack of understanding why the post had been removed, and around a third (29\%) were frustrated by the removal of their post. These results indicate that frustration and lack of acceptance are common among users whose content has been moderated.

Previous research sheds light on how to address perceptions of unfairness in moderation decisions. First, community participation, in which community members develop rules and explanations, appears to be a feasible route \cite{kou_mediating_2020}. Second, investing time and effort in explaining the decisions to banned users can be beneficial in avoiding perceptions of exclusion \cite{jhaver_does_2019} and helping to realise the civic role of social media platforms in enabling open and inclusive public discourse \cite{myers_west_censored_2018}. Ideally, explanations of moderation decisions instruct users about the social norms of the platform, which guide them to behaviours that the platform owner perceives productive \cite{jhaver_does_2019}. This educative model of content moderation (as opposed to punitive) is believed to promote healthy online communities \cite{myers_west_censored_2018}. A previous study found that whether explanations of moderation decisions were generated by human admins or bots did not have a significant difference, but in both cases removal explanations were effective in guiding user behaviour \cite{jhaver_does_2019}. The use of explanations is also supported by findings showing that users who are knowledgeable about a platform’s community guidelines or were explained the reasons for removal are more likely to perceive the moderation decision as fair \cite{jhaver_did_2019}. However, it is unclear how many users on FB and other dominant platforms actually read and understand the community guidelines. 

Third, designing contestability, i.e., giving users means to influence how content moderation decisions are made (and appeal about decisions they consider unfair) is considered important \cite{vaccaro_contestability_2021}. Contestability design is needed because content moderation systems must ``navigate inherently normatively contestable boundaries'' \cite{binns_like_2017} (p. 405), meaning that some of their decisions are contestable and subjective (partly derived from norms and assessments provided by human annotators of training data).
Fourth, at worst, content moderation may be applied unequally for different groups of users, resulting in disparate outcomes for users based on their gender, race, or political orientation \cite{haimson_disproportionate_2021}. Research indicates that fringe user groups like political conservatives, transgender users, and African-American users experience content and account moderation more frequently than other users \cite{haimson_disproportionate_2021}. This implies that there is a need for cultural sensibility in analysing and implementing moderation policies.


\subsection{The Need for Understanding Content Moderation Better}

There are various theories about the interaction between platforms and their users. We examine some of these, and the potential ramifications that limiting users' freedom of speech can have when interpreted using these theories.

\citet{salminen_platform_2018} perceive the relationship between users and social media platforms as a form of  a\textit{social contract}, in which both parties commit to certain informal rules not present in the official terms of service. In the classic social contract theory, which originates from French philosophers of the Enlightenment era \cite{rousseau_social_1999}, the relationship between the state and its citizens is not defined as a master--subject ordinance, but as two parties that willingly give away some rights in exchange for others. In the context of platforms, for example, users allow the platform to use their data for advertising in exchange for free access and usage \cite{salminen_platform_2018}. Platforms, in turn, commit to not divulge personally identifiable or otherwise sensitive information and to treat their users equally. Limiting a user's freedom of expression can therefore be interpreted by the users as a breach of the social contract, which would undermine the platform's legitimacy to moderate its users. If there is a low switching cost between platforms, users might abandon the platform that is perceived as unfair \cite{rochet_platform_2003}, or carry out acts of ``rebellion'' (e.g., coordinated downvoting campaigns in app marketplaces).

Noelle-Neumann \cite{noelle-neumann_spiral_1974} suggested the \textit{spiral of silence theory} in her seminal article of 1974, according to which society excludes or isolates members based on members' deviant opinions. The existence of this social threat results in carrying out self-moderation (i.e., voluntary limitation of freedom of expression) due to the fear of exclusion. This fear emerges from unconsciously observing acceptable and unacceptable behaviours--for example, what type of content is promoted or removed by a social media platform's algorithm. This theory is similar to signalling theory\footnote{``Signalling is the idea that one party credibly conveys some information about itself to another party'' \cite{morris1987signalling}.}, implying that community guidelines and moderation practices signal socially (dis)allowed behaviours to users. From a design point of view, this is vital, as each ranking choice made by the newsfeed algorithm shapes not only the worldview of a user \cite{pesce_last_2017} but also their conception of \textit{what speech} is allowed on the platform. In other words, the newsfeed algorithm is an instrument for moral judgement, without this being explicitly hard-coded into its operating logic. These automatic choices constitute an implicit algorithmic bias \cite{johnson_algorithmic_2020}. Because filtering information about what others are thinking or doing \cite{margetts_political_2017} influences a person's political decisions, such as voting behaviour \cite{bischoff_social_2013}, the suppression of certain opinions in social media can have second-order ramifications in society at large.

Moreover, the existence of bias would not be met passively by the users. According to the theory of \textit{group polarisation} \cite{myers_group_1976}, users are divided into groups based on their differences of opinion. This is already visible in the American political scene, where conservatives who perceive being unfairly censored by social media platforms, seek to create alternative platforms \cite{abril_conservative_2021}. The result could be that Platform A develops into an environment where only a subset of political spectrum is displayed, while Platform B favours the opposite views. The negative consequence of this is that users of Platform A would now be less aware of the opinions of the users in Platform B, and \textit{vice versa}. This effect is often referred to as filter bubbles \cite{pariser_filter_2011} or echo chambers \cite{del_vicario_echo_2016}, implying that the presence of like-minded users reinforces the group's dominant belief system. While such filter bubbles may result in less conflict in the short term (because users with opposing views have less probability of interaction and therefore less chance of a quarrel), in the long term, the theory of group polarisation suggests that, with no exposure to alternative opinions, the dominant opinions of each group become entrenched and stronger than if there were a diversity of opinions. This entrenchment can manifest itself in more extreme behaviours than would otherwise occur. Polarisation can also affect users of the middle opinion towards either extreme of a binary choice (e.g., ``Brexit or no''; one is either for or against \cite{alvim_toward_2019}). Hence, the compartmentalisation of group-specific social media platforms cannot be viewed as a desirable course for society.

Furthermore, it is unclear how well the design of social media platforms' content moderation policies accounts for \textit{cultural sensibility theory}, which refers to considering cultural variability in design choices of technology \cite{hakkila_design_2020}. In the worst case, there can be structural bias against a certain culture due to misinterpreting the meaning of certain hashtags and keywords used \cite{lamarre_hashtag_2017}.
ML models for moderation are often trained on English-language datasets, which means that content in different languages (e.g., Arabic) and cultural contexts may be at a structural disadvantage when it comes to filtering content. 
The existence of cultural insensitivity can contribute to the worsening of the global digital divide \cite{warschauer_reconceptualizing_2002}, which global platforms such as FB desire to avoid, according to the company's mission statement\footnote{``Founded in 2004, Facebook's mission is to give people the power to build community and bring the world closer together. People use Facebook to stay connected with friends and family, to discover what is going on in the world, and to share and express what matters to them.'' \cite{facebook_facebook_2021}}. Such effects can be observed in the ratings given by users to the platform in reputation systems, e.g., when users who felt that FB is removing their content unfairly gave FB one-star ratings \textit{en masse}, resulting in FB's decreasing ratings in app marketplaces~\cite{FB_rating_1,FB_rating_2}. Since these app marketplace ratings matter for a platform's reputation \cite{fan_reputation_2016} and affect trust between users and platforms \cite{wang_understanding_2014}, perceptions of unfairness can have dire consequences for social media companies.

In conclusion, based on prior research and theoretical work, unfair treatment of users on social media platforms can have wide repercussions not only for users but also for multiple other stakeholders, such as platforms (undermining their legitimacy), society at large (increasing conflict, group polarisation, and filter bubbles), and researchers. Researchers are affected because systematic suppression of political beliefs (e.g., by banning certain keywords or hashtags \cite{lamarre_hashtag_2017}) can result in biased data collection. A hypothetical example would be that if FB would ban pro-Palestine hashtags but not pro-Israel hashtags, social scientists using these hashtags to collect data about public opinion would now conclude that the public supports the Israeli cause and hardly anyone supports the Palestinian cause. If such studies were used in the process of political decision making, there would be a risk of cascading errors due to false conclusions.  
Thus, free expression--or at least politically equal moderation--is a prerequisite for unbiased data collection. 

This, and other risks of unfair treatment of users based on their political views, underline the importance of studying platforms' moderation process and user perceptions of its fairness. It is not evident that platform developers are aware of the full extent of their design power in this context \cite{johnson_algorithmic_2020}, 
which is why research on fairness experiences on social media is highly recommended.  

\section{Data Collection of FB-Moderated Posts}
\label{sec:data}

In this section, we describe our process for collecting posts that have been moderated and removed by FB. Then, we discuss some statistics on the collected data and its nature.

\subsection{Ethical Consideration}

The Palestine-Israel conflict is one of the most sensitive political topics around the world. Taking a stance towards one side over the other is usually associated by shame from supporters of the other side. This can be illustrated by the incident in January 2022 when the British actress Emma Watson posted an image on Instagram showing a photograph of a pro-Palestinian protest. This sparked accusations of antisemitism from supporters of Israel, including Israeli officials~\cite{emmaattack}. Similarly, on the other side, when the candidate for 2021 NYC mayoral election, Andrew Yang, tweeted a statement during the conflict in May 2021 supporting Israel and attacking Hamas, a backlash came accusing him of ignoring the Palestinian victims. This led him later to make another statement acknowledging the suffering on both sides~\cite{yangattack}. These examples demonstrate the sensitivity of this polarised topic.

During our study of this sensitive topic, we took several measures to ensure ethical research and avoid bias in our data collection, annotation, and analysis. This includes, as detailed in the following subsections, avoiding the collection of any posts that might contain any information about the identity of author; we also give clear instructions to annotators to label posts independent from their own position and we applied quality control to ensure the objectivity of the labelling process. We also make it clear to annotators that some posts might contain hate-speech and explicit content before taking the job. The identities of all participants in this study, including survey participants and post annotators, are fully anonymous. Moreover, we have all our analysis steps well-planed prior to data collection to avoid any possible bias in the analysis.

In terms of the authors of the current research, the research team consists of two Arabs who sympathize with the Palestinians and a Caucasian who is neutral towards the conflict. All researchers in the research team are senior researchers who published several works in the HCI community. 
The team made every effort to uphold scientific objectivity throughout the research process.

\subsection{Collecting Deleted Posts}

The 2021 unrest between Palestine and Israel began in April, but media attention soared in May~\cite{dwoskin_facebooks_2021}. On 23 May 2021, we released a survey form\footnote{Survey form can be found at: \url{https://bit.ly/3xVXCi9}} and invited Arab users who were moderated on FB to participate by providing their posts that FB has removed. This took place while the campaign of down-rating FB's app started to be effective and grab media attention. The form instructed the participants on how to extract their deleted posts and any external links (e.g., photos and videos) that were associated with the posts\footnote{In FB, deleted posts can be viewed by selecting: ``Posts/Account Warning/Restrictions/See Why'' under profile information.} and how to submit the posts for the study.

The form included a consent section that informed the participants about the purpose of collecting these posts, which is investigating the claims of FB bias and analysing the topics that are moderated by FB. Participants were clearly instructed that: (1) they must provide only posts deleted by FB that led to restricting the usage of their account; (2) the posts must be from their own individual account, (3) they need to hide any personal information in the post that might reveal the identity of any user; and (4) the posts would be made public for research purposes. The participants had to agree on these four points before proceeding.

In the form, users were asked to provide information about their deleted posts, including (1) the date of the post, (2) the date when the post was deleted by FB, (3) the nature of the post (text, image/video, or link), (4) the text of the post as accurate as possible (without any personal information), (5) a link to the media they shared, if any, and (6) a description to the media they shared if no link to a similar media is available. In addition, we asked the participants about (7) the restriction applied to their account, (8) if they appealed the decision, and (9) the outcome of the appeal. We also asked them to optionally provide their demographic information, including their (10) gender, (11) age group, (12) country, and (13) the total number of friends and followers they have on their FB account. Finally, we ask whether (14) the participant thinks that FB is biased against certain groups or people and how often they experience bias.

The survey form was designed to record only one post at a time, and participants were instructed to fill it multiple times for multiple posts if needed.
The form link was published on the FB account of one of the authors that had around 10,000 followers (majority Arabs), and followers were invited to participate if they had experienced any moderation from FB in the past. In addition, the followers were encouraged to share the form within their social networks and invite others to participate. The link to the survey was shared more than 1,300 times on FB, and within two weeks, we received 588 responses.

\subsection{Data Cleaning and Verification}
The collected 588 responses were carefully checked for validity. We observed that some of the responses submitted did not include the post as is, but just a description of the post content (e.g., ``\textit{The post was about supporting Palestine}''). In addition, some posts which were described as photos/links did not provide a link to a similar photo or a clear description of the content in the photo. Thus, we filtered out these posts. This process led to a collection of 448 posts (76\% of the 588 responses) that included verbatim text and full details.

For those 448 posts, we manually coded the topic in each, which was done collaboratively among the researchers. 
Two researchers reviewed the posts and applied thematic analysis to determine the specific topic discussed in each post. The posts were placed in a shared spreadsheet, one researcher went through and sorted the posts into topics. They then met and the second researcher who read through all the posts and adjusted the topics discussing with the first researcher as they went. The result was a set of specific topics that both researchers agreed upon. The initial number they assigned the posts to was 47 topics. The two researchers then met again to map these specific topics into general themes to facilitate the analysis. This process led to mapping the 47 specific topics into seven main themes: Palestine, Israel, Palestine resistance, Jews, Religious groups, LGBTQ, and others (see Table~\ref{tab:agreement}).
As we expected, most of the topics were related to the Israel-Palestine conflict.

\begin{table}[t]
\centering
\small
\begin{tabular}{p{1.2cm}p{6.5cm}}
\hline
\textbf{Theme} & \textbf{Description} \\
\hline
Palestine & Posts showing support to Palestine and Palestinians\\
\hline
Israel & Posts about Israel, its cities, and Israelis\\
\hline
Palestine resistance & Posts supporting Palestine resistance, Hamas, Qassam, or their figures\\
\hline
Jews & Posts about Jews or Zionists (usually against them)\\
\hline
Religious groups & Posts about Muslims, Christians, Sunni, Shia, or Atheists\\
\hline
LGBTQ & Posts about the LGBTQ community\\
\hline
Others & Posts about other countries, suicide, men, women, and others\\
\hline
\end{tabular}
\caption{The main seven themes (topics) discussed in the collected 448 posts.}
\label{tab:themes}
\end{table}

As a primary verification step to verify that these collected posts are likely deleted by FB and not invented by respondents, we created a FB account and posted a random set of 10 posts of these collected ones. We noticed that FB banned the account the following day, which supports the notion that these posts contained content that FB algorithms classify as being against their community guidelines.

\subsection{Statistics on the Collected Posts}

\subsubsection{\textbf{Participant demographics}}
\begin{figure*}
\centering
\includegraphics[width=0.8\linewidth]{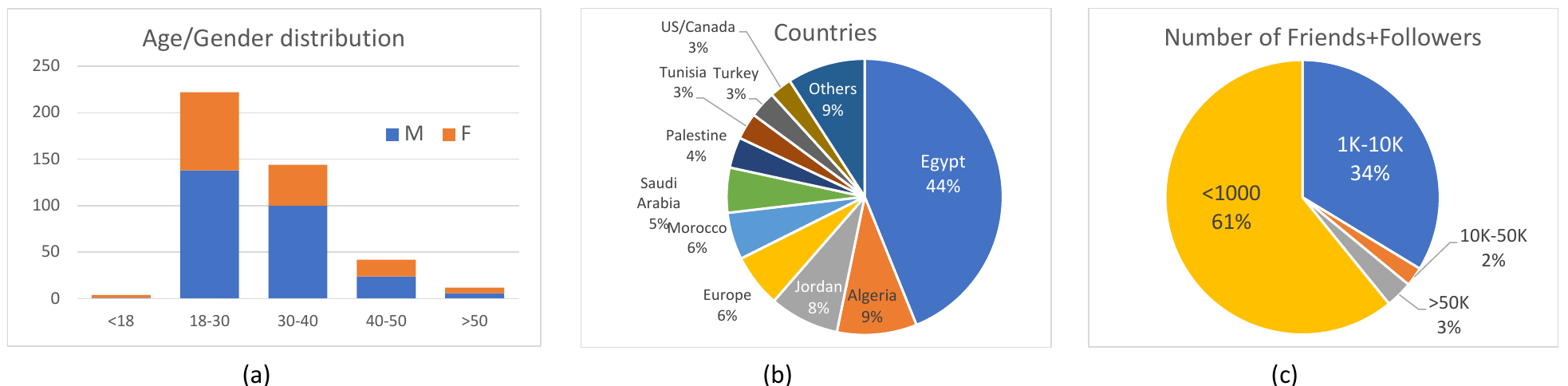}
\caption{Demographics and account reach of the authors of the 448 deleted Facebook posts}
\label{fig:demographics}
\end{figure*}

Figure~\ref{fig:demographics} shows information about the participants whose posts were included in our analysis. Figure~\ref{fig:demographics}(a) reports the age and gender of the participants who provided this information. As shown, most of the participants were male (63.5\%) and between 18-40 years old (86\%). It was interesting to find that four posts in our collection were from minors (<18 years of age). Regarding the location of our participants, as shown in Figure~\ref{fig:demographics}(b), Egypt was the most common location (44\%), which could be due to the large population of Egyptians present on FB and in the network of the author posting the survey form. Other locations included Algeria (9\%), Jordan (8\%), Morocco (6\%), and Saudi Arabia (5\%). We also received posts from participants in Europe (6\%), United States (US), and Canada (3\%). One of the limitations of our sample is that we have only 4\% of the responses from the participants in Palestine. However, this might be seen as reasonable given the smaller population size of Palestine compared to other Arab countries. Finally, Figure~\ref{fig:demographics}(c) shows that most participants have fewer than 1000 friends and followers (61\%), while 34\% have 1-10K friends and followers. We also received 10 (2\%) and 14 (3\%) responses from participants who have 10-50K and over 50K followers, respectively. Information about sex, age, country, and number of friends was available for 95\%, 96\%, 93\%, and 99\% of all participants, respectively.

\subsubsection{\textbf{Deleted posts and restrictions applied}}
\begin{figure*}
\centering
\includegraphics[width=0.8\linewidth]{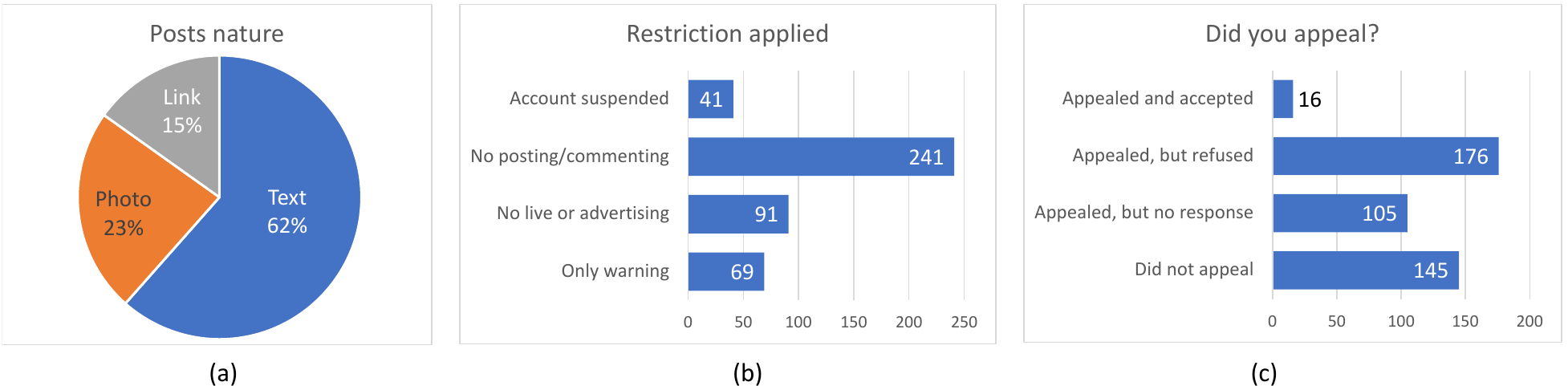}
\caption{Statistics on the nature of the posts (a), restriction applied to the account (b), and appeal on restriction for the 448 posts (c).}
\label{fig:posts}
\end{figure*}

Figure~\ref{fig:posts} shows statistics on the nature of posts and the restrictions applied to the accounts published them and if there were any follow-up appeal on the restriction and the subsequent decisions. As shown in Figure~\ref{fig:posts}(a), most of the posts we collected (62\%, n=272) were just textual status updates. In contrast, 23\% and 15\% of the posts were photos and external links along with textual commenting. The posts' text length varies considerably, between only one word (in a few cases when the post contained a photo or a link) to over 100 words. The median length of the textual content of the posts is 11 words. 

Regarding the type of restriction the account holders received when FB deleted their post, as shown in Figure~\ref{fig:posts}(b), most of them (55\%) were banned from posting or commenting on FB for a given period of time (most of them for 24 hours or a few days), 21\% were banned from advertising on FB or getting live for a month, 16\% were only warned, and 9\% got their FB account completely suspended. When we asked our participants if they appealed to the restriction applied and how FB responded, as shown in Figure~\ref{fig:posts}(c), the majority of the accounts have appealed (67\%). However, those appeals were either refused or ignored by FB. Only 16 respondents (3.6\%) appealed and succeeded in having the restriction removed from their account.

For our question in the survey form about FB bias, 93\% of the participants described FB as ``highly biased'' and 5.4\% as ``sometimes biased'', while less than 0.5\% see FB ``mostly fair'', and the remaining 1\% selected ``I don't know''. This result is not surprising, since our sample of participants is skewed toward those who perceive FB is treating them unfairly, as we only surveyed users who got their posts deleted by FB. Although we may not generalise the results to the general population of FB users or Arab users in FB, it is evident that the vast majority of the participants whose posts were moderated by FB considered the decision as biased and unfair.

\subsubsection{\textbf{Topics of the deleted posts}}
\begin{figure}
\centering
\includegraphics[width=0.7\linewidth]{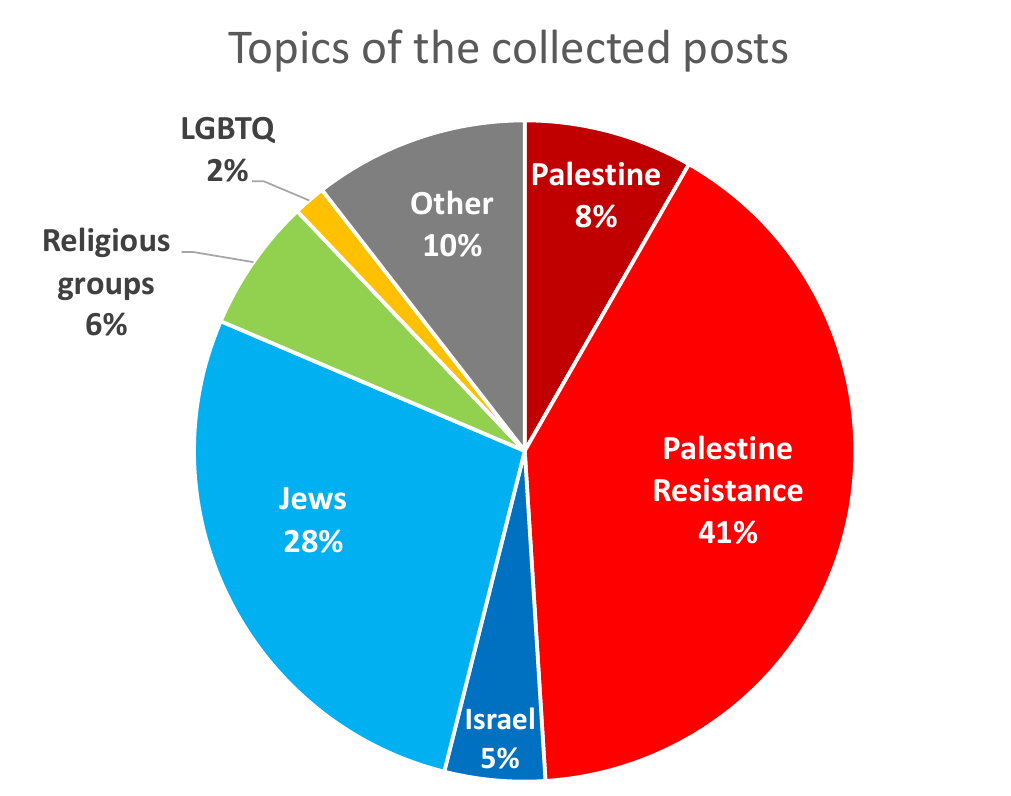}
\caption{Distribution of the topics discussed in the collected 448 posts.}
\label{fig:topics}
\end{figure}

As mentioned earlier, we manually labelled the 448 posts we collected according to the topic discussed in each of them. As expected, most of the posts were related to the Israeli-Palestine conflict (82\%) and the remaining 18\% discussed other various topics. Figure~\ref{fig:topics} shows the distribution of these topics that can be classified as follows: 41\% of the posts were about the Palestinian resistance and its figures, including militant and political groups such as Hamas, Al-Qassam, and figures such as the founder of Hamas, Sheikh Ahmed Yassin. The second largest topic, which covered 28\% of the posts, was discussing Jews and subgroups, such as Zionism. Furthermore, 8\% and 5\% of the posts were discussing Palestine and Israel, respectively. The remaining 18\% of the posts covered various topics, including religious groups in the Middle East (6\%) such as Sunni, Shia, and atheists; LGBTQ (2\%), and other various topics (10\%) covering specific countries, nationalities, genders, and others.

In the following, we describe our methodology to validate if these 448 posts deserved to be moderated by FB according to the platform's community standards and according to Arab users' opinion.


\section{Investigating the Implementation of the FBCS}
\label{sec:method}

To address our RQs, we annotated our collected posts, presenting the rationale given by the FBCS, and seeing if the annotators agree with these rationales. In this section, we describe our methodology for annotating our data and the analysis process. First, we discuss the FBCS and how we included them in our annotation process. 

\subsection{Facebook Community Standards (FBCS)}
The FBCS\footnote{\url{https://www.facebook.com/communitystandards/}} contain a detailed list of behaviour and content regulations that should be taken into account when sharing content on FB. As stated by FB, these standards are set \textit{``to ensure that everyone's voice is valued''}, and they \textit{``include different views and beliefs, especially from people and communities that could otherwise be overlooked or marginalised''}. The FBCS also state, \textit{``Facebook company wants people to be able to talk openly about the issues that matter to them, even if some may disagree or find them objectionable''}. In the following analysis, we put these statements to test and validate their implementation among Arab users.

The FBCS cover five main aspects: (1) Violence and criminal behaviour, which include spreading violence or showing support to dangerous individuals/organisation; (2) Safety, which includes self-harm, human exploitation or abuse, harassment, and privacy violation; (3) Objectionable content, such as hate-speech and graphical/sexual content; (4) Integrity and Authenticity, including spamming and misinformation; and (5) Intellectual property (IP) violation.
Table~\ref{tab:FBCS_list} shows each of these aspects along with the sub-aspects of each with their definition.

For our annotation task, we translated those aspects and their sub-aspects (24 sub-aspects in total for the five aspects) into Arabic to be used as guidelines for our annotators.

\begin{table*}[t]
    \centering
    \footnotesize
    \begin{tabular}{p{1.5cm}p{15cm}}
    \toprule
       FBCS Aspect  & List of sub-aspects \\
       \midrule
       Violence & - \textit{Violence and Incitement}: Incitement to or facilitation of serious violence, such as a real risk of physical harm or direct threats to public safety.\\
        and Criminal Behaviour & - \textit{Dangerous Individuals and Organizations}: Supporting or praising groups or individuals engaged in terrorist activity, organized hatred, mass murder, multiple murders, human trafficking, organized violence or criminal activity\\
       & - \textit{Coordinating Harm and Propaganda of Crime}: Facilitating, organizing, promoting or acknowledging criminal or harmful activities targeting people, businesses, property or animals\\
       & - \textit{Regulated Goods and Regulations}: Buying, selling or trafficking non-medical drugs, narcotics and marijuana. Also buying, selling, gifting, exchanging and transporting firearms and ammunition\\
       & - \textit{Fraud and Deception}: Deceiving or exploiting others for money or property\\
       \midrule
       Safety & - \textit{Suicide and self-harm}: Celebrating or promoting suicide or self-harm intentionally or unintentionally\\
       & - \textit{Child sexual exploitation, abuse and nudity}: Sexually exploiting or endangering children\\
       & - \textit{Adult sexual exploitation}: Depicting, threatening or promoting sexual violence, sexual assault or sexual exploitation\\
       & - \textit{Bullying and harassment}: Making threats, issuing personally identifiable information or sending threatening messages and making unwanted harmful communications\\
       & - \textit{Human exploitation}: Facilitating or coordinating the exploitation of humans, including human trafficking, such as depriving someone of their liberty for profit or forcing them to engage in commercial sex or labor against their will\\
       & - \textit{Privacy and image privacy rights violations}: Sharing, displaying or soliciting personally identifiable information or other private information that could lead to physical or financial harm, including financial, housing and medical information, as well as information from illegal sources\\
       \midrule
       Objectionable Content & - \textit{Hate speech}: Direct attacks (insults, contempt, incitement, etc.) against people on the basis of: race, national origin, disability, religious affiliation, social class, sexual orientation, gender, gender identity, serious illness\\
       & - \textit{Violent and graphic content}: Glorifying violence or celebrating the suffering or humiliation of others\\
       & - \textit{Sexual exploitation of adults}: Depicting, threatening or promoting sexual violence, sexual assault or sexual exploitation\\
       & - \textit{Adult nudity and sexual activity}: Displaying nudity or sexual activity\\
       & - \textit{Sexual solicitation}: Facilitating, encouraging or coordinating sexual encounters or commercial sexual services between adults such as prostitution\\
       \midrule
       Integrity and & - \textit{Account Safety and Identity Authenticity}: Impersonation and Identity Misrepresentation\\
       Authenticity & - \textit{Spam}: Tricking or misleading users to increase viewership\\
       & - \textit{Cybersecurity}: Attempts to collect sensitive user information or gain unauthorized access\\
       & - \textit{Inauthentic Behavior}: Using fake accounts or artificially increasing the popularity of content\\
       & - \textit{Fake News}\\
       & - \textit{Manipulated Media}: Manipulating media (images, audio, or video) for the purpose of misleading\\
       & - \textit{Memorialization}: Attempts to log in and fraudulent activities of a person after their death\\
       \midrule
       IP Violation & - \textit{Intellectual Property}: Failure to respect the copyrights, trademarks, and other legal rights of others\\
       \bottomrule
    \end{tabular}
    \caption{The list of five aspects covered by the FBCS and the corresponding sub-aspects of each with their definition. For the data annotation process, the translated version of each aspect was shown separately to annotators who were asked to select the sub-aspects that a given post might be violating.}
    \label{tab:FBCS_list}
\end{table*}

\subsection{Data Annotation and Quality Control}

We created an annotation job on the Appen crowdsourcing platform\footnote{\url{www.appen.com}} to label the 448 posts we collected as violating any of the standards we translated from the FBCS. To receive objective assessments, we presented this task as a general annotation job without mentioning any relation to FB. However, due to the expected sensitivity of posts, we made it clear to annotators that the job may contain explicit content, which is flagged by Appen platform to annotators before taking the job.
We paid \$15 per hour of work to comply with the minimum wage rate in the US. The annotation process was performed and completed during the first half of November 2021.

Annotators were asked to read the guidelines carefully and given clear instructions to check if each post violates any of the five main aspects of community standards. 
Annotators were asked about each aspect separately to indicate if the post is violating it or not by selecting the exact sub-aspect violated. For a question about a given aspect, all the sub-aspects and their definition (as listed in Table \ref{tab:FBCS_list}) were listed to the annotator to select the relevant violated aspect, if any. Definitions were listed to ensure high-attention and avoid any confusion by annotators. A ``non-of-the-above'' option was available with the questions on each aspect to be selected when none of the sub-aspects apply to the post.
After the five questions about each aspect, we added a final question asking the annotator if they think that the post should be removed according to their own personal opinion regardless of the guidelines. We use this question to address RQ2.
Posts with a photo or link were constructed in the annotation job to include the photo and link, which ensures that the annotators see the post in a format that corresponds to how the post was visible on FB.

We set each post to be assessed by 10 annotators from Arab countries to obtain an adequate number of assessments for the analysis. We specifically focus on Arab users in the Appen task because the posts were in Arabic. Also, this is compatible with our RQs that focus on the views of Arab FB users. 

To ensure that the annotators paid full attention to the annotation job and to avoid personal bias in the annotation, we implemented the following five steps:

\begin{enumerate}
\item We did not inform the annotators that the posts were from FB users or they had been removed, but we indicated to them that we want to make sure that these posts do not violate certain community standards. 

\item We set the job to be done only by annotators who have the highest quality rating based on their history with Appen.

\item We emphasised to the annotators that judgments should be based on the standards described in the guidelines, regardless of their personal opinion.

\item We manually crafted 50 additional quality control (QC) posts for the purpose of evaluating the quality of the annotators' work. Of these, 25 clearly violate one of the standards (e.g., containing racism, pornography, or incitement to violence) and the other 25 clearly do not violate any of the standards. These posts were randomly inserted into the original posts as part of the quality control process within Appen, and the annotation job was set to exclude the judgements of any annotator who achieves less than 80\% accuracy on the QC posts. The purpose of this procedure was to control the quality and to verify the attention of the annotators throughout their work on this task.

\item Rejected annotators were replaced by additional workers until we had ten quality judgments for each post.
\end{enumerate}

The annotation job was performed by 106 different Arab annotators from six different countries (Egypt, Algeria, Palestine, Saudi Arabia, and Tunisia), showing a diversity of Arabs who performed the job. The maximum number of posts annotated by a single annotator was 48. The average performance of those 106 annotators who passed our quality control process was 94\% on our QC posts, which confirms their high performance in performing the task.

As a result, for each of the 448 posts in our collection, we obtained 10 judgements from different annotators stating if the post is violating any of the FBCS with specification to the aspect violated; in addition, we obtained another 10 judgements based on the annotators' personal opinions of whether the post should be removed. 
The posts, their metadata, and the full list of judgments are made public and can be downloaded\footnote{\url{https://osf.io/eupqm/}} for research purposes.


\section{Results and Analysis}

In this section, we analyse the results of the annotated data to answer our RQs.

\subsection{RQ1 and RQ2: Did the Posts Violate the FBCS? How Did Arab Users Agree with FB's Moderation?}

Here, we analyse the assessments and the extent to which annotators agreed that posts should be deleted or kept on two occasions--once based on the FBCS (RQ1) and once based on the annotators' own opinions (RQ2).

Initially, we checked the overall percentage of judgments that indicated that a post should be removed across all the 448 posts. Of the 4480 judgements we collected (10 judgments $\times$ 448 posts), only 40.6\% indicated that the posts violate the FBCS, while the remaining 59.4\% found the posts did not violate any of the aspects of the FBCS, see Table~\ref{tab:agreement}. This result indicates that the implementation of the FBCS is misapplied on most of the deleted Arabic posts we collected.
This percentage becomes even larger when we consider the personal opinion of the annotators instead of the FBCS, where 71\% of the judgements considered that the posts have nothing that requires its deletion. This indicates that around 12\% of the posts
that the annotators found violating the standards, the annotators themselves showed that they should not be removed based on their own opinions.
This initial result might explain the disappointment of some Arabs with FB and their view of it as a biased platform. 

We also measured the agreement between annotators with respect to both methods of annotation using the Fleiss kappa~\cite{landis1977measurement}. The agreement among the annotators when judging the violation of posts in the FBCS was $\kappa$=0.522, which indicates a moderate agreement, suggesting that the FBCS helps the annotators identify what should be removed, though perfect agreement remains elusive even with explicit guidelines. 
This agreement was $\kappa$=0.366 when the annotators considered only their personal opinion, which is a fair agreement~\cite{landis1977measurement}.
These kappa scores reveal important insights about content moderation in cross-cultural contexts. The moderate agreement when using FBCS ($\kappa$=0.522) suggests that while the guidelines provide some common ground for decision-making, there remains significant room for interpretation even within the same cultural group. The lower agreement score for personal opinions ($\kappa$=0.366) is logical since no clear guidelines are used, but it also highlights how cultural values, while shared, still lead to diverse individual interpretations of content appropriateness.
These findings suggest that achieving consistent content moderation decisions is challenging even within a single cultural context, let alone across different cultural groups. This inherent subjectivity in content interpretation underscores the complexity of developing and implementing global moderation standards that can be consistently applied across diverse cultural contexts.

\begin{figure}
\centering
\includegraphics[width=\linewidth]{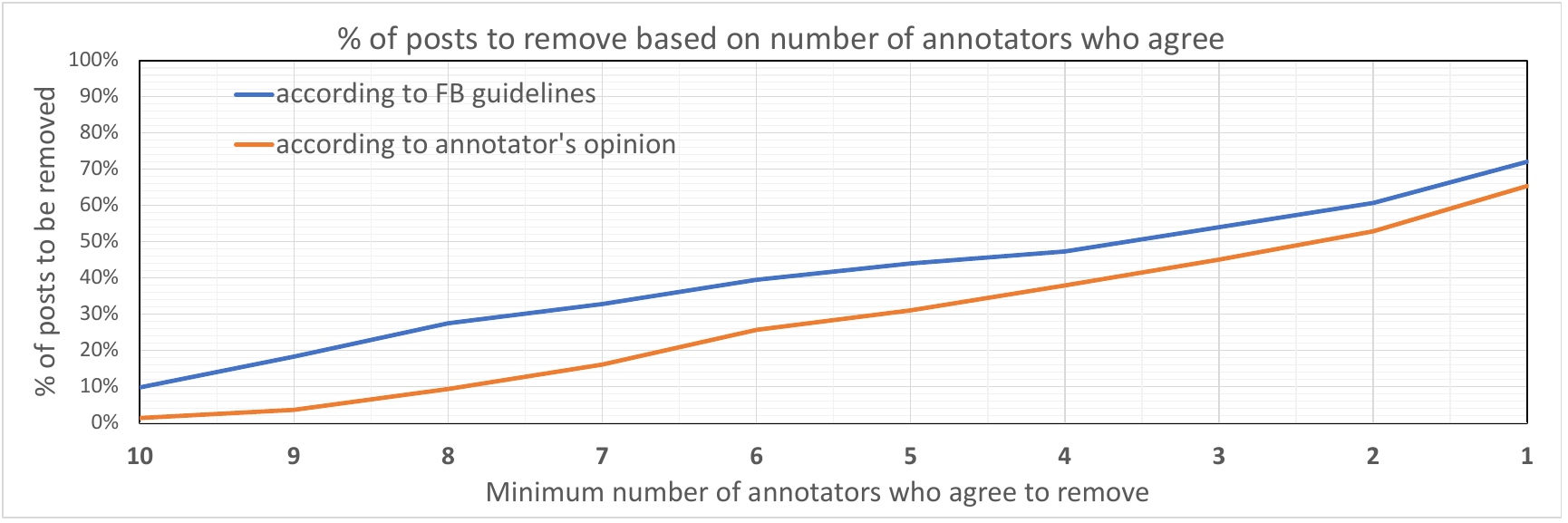}
\caption{Percentage of posts that should be removed based on the minimum number of annotators who agree that they violate the FBCS or according to their personal opinion.}
\label{fig:removebyN}
\end{figure}

Taking into account that each post was judged by 10 annotators, Figure~\ref{fig:removebyN} shows the number of posts that N/10 annotators agree that they should be removed, once according to the FBCS and once according to their own opinion. Only 10\% (n=44) of the posts have 10/10 of the annotators agreeing that they violate the FBCS. However, when considering the annotators' opinion, only 1.3\% (n=6) of the posts received 10/10 because they are improper and should be removed. On the other hand, for 28\% (n=125) of the posts, none of the 10 annotators (0/10) found that they violate the FBCS in any way, while 35\% (n=155) of the posts annotators think they should not be removed according to their opinion.

If we consider the majority votes on posts, the proportion of posts where 7 or more annotators agreed that they did not violate the FBCS was 53\% (n=236). In contrast, only 33\% (n=147) of the posts got 7 or more votes that they violate the standards and should be removed, which leaves the remaining 15\% of the posts controversial (votes were 4-6/10).
However, if we consider the annotators' personal opinions instead, these percentages lean more towards not deleting the posts. The percentage of posts where 7 or more annotators agreed that they should not be deleted reaches 62\% (n=278), while the percentage of what should actually be deleted--if 7 or more annotators agree--is only 16\% (n=72). 

The above results imply that most of the deleted posts by FB in our collection were a result of incorrect decisions as assessed by the Arab annotators, either evaluated based on the FBCS or based on the annotators' personal opinions. These results address both our RQ1 and RQ2. In the following, we apply a more quantitative and qualitative analysis by examining patterns of disagreement among annotators through analysing the topics of the posts that were seen as misclassified by FB for violating its FBCS.

\begin{table}
\centering
\begin{tabular}{l c c} 
Question & YES & NO\\
\hline
Does the post violate any of the FB guidelines? &  40.6\% & 59.4\% \\
Do you think the post should be removed?
 & 28.8\% & 71.2\% \\ 
\hline
\end{tabular}
\caption{Overall percentage of annotators who agree that the post should be removed across all posts}
\label{tab:agreement}
\end{table}

\subsection{RQ3: How Does the Topic of Moderated Content Affect Users' (Dis)agreement with the Moderation?}

\begin{figure*}
\centering
\includegraphics[width=0.85\linewidth]{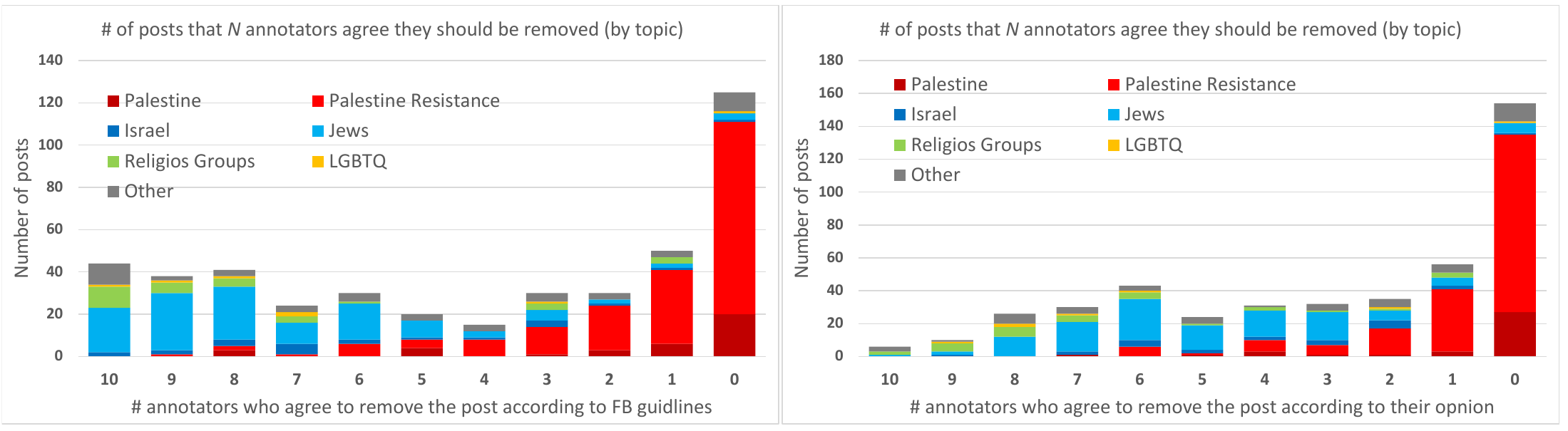}
\caption{Topics of the posts that annotators judged that they should or not be removed based on the FBCS (a) and their personal opinion (b).}
\label{fig:removebytopics}
\end{figure*}

Figure~\ref{fig:removebytopics} shows the number of posts for each N annotators out of 10 who agreed that (a) they violate the FBCS, and (b) they should be removed according to their personal opinion. Both have the topics marked on the chart. As shown in both Figures~\ref{fig:removebytopics} (a) and (b), topics related to Jews, Israel, Religious groups, and LGBTQ are more likely to receive more votes stating that they violate the FBCS and should also be deleted according to the annotators' opinions. The posts related to these topics usually contain hate speech against certain groups of people, and thus there is more agreement that the content is improper and should be removed. However, topics related to Palestine and Palestine resistance are more likely to be seen to be not violating any of the FBCS, and also most annotators believe such posts should not be removed according to their personal opinion. The posts related to these topics are mainly showing support for Palestinians and their resistance.

These results address our RQ3 and can be seen as highly interesting, since they highlight where the misalignment between FB's implementation of the FBCS and how Arab users perceive these standards. It seems that Arab annotators easily spot when hate speech occurs in a given post against various groups of people, including Israelis who might be seen as enemies in the Arab world, where 88\% of Arabs do not recognise the state of Israel, according to the 2021 survey by \citet{almasri2021assessing} which covered 14 different Arabic counties, including the six countries our annotators come from. Interestingly, the annotators managed to apply the guidelines and identify the violations of the FBCS even when the standards were against their own opinion, as can be seen from the difference in the distribution of topics in Figures~\ref{fig:removebytopics} (a) and (b). For other topics related to supporting Palestine and the Palestinian Resistance, it becomes clear that there is a large disagreement between the annotators and FB decision on deleting the posts. There might be several reasons for this disagreement with FB's decision. Is it because FB's algorithms might be biased and made a wrong decision by deleting these posts? Or is there a misunderstanding of the FBCS by the Arab annotators that these posts are indeed violating the FBCS but this is not observed by the annotators?

In the following, we discuss some examples of these posts to qualitatively better understand the possible reasons for this disagreement about FB's decisions to moderate the posts.

Table~\ref{tab:examples} shows 13 examples of the posts in our collection along with the votes on each by the annotators for violating the FBCS. As shown, examples with low values of N/10 do not contain hate speech but rather statements of support for Palestine, the resistance of Palestinians, or its human figures. There are also examples of statements supporting the education of women that may be incorrectly classified by FB's algorithm as hate speech against women (see Example 3), and some facts about LGBTQ in the Arab culture (see Example 5). In Table~\ref{tab:examples}, N/10 has larger values, involving examples of clear hate speech against certain groups, such as Jews, Arab atheists, and Indians, which all received more agreement from the annotators in terms of violating the FBCS.

These examples corroborate our earlier finding that Arab annotators agree with FB's decisions of removing posts that contain hate speech. However, the annotators disagree on examples that contain no hate speech and are likely to be seen by FB as violating other aspects of the FBCS.

\begin{table}[t]
\centering
\footnotesize
\begin{tabular}{p{6.7cm}cc}
\toprule
& \multicolumn{2}{c}{voted remove} \\
\cmidrule(lr){2-3}
Post (translated*) & FBCS &  PO\\
\midrule
Palestine - from the sea to the river - is Arabic and Islamic & 0 & 0\\
May God have mercy on you, Sheikh Ahmed Yassin .. Palestine wins & 0 & 0\\ 
Ignorant Muslim woman means ignorant nation. Seek knowledge ... & 1 & 1 \\
Down with the Israeli aggression on Palestine \#Support\_the\_resistance & 2 & 2\\
There're organisations support homosexuals, even in our Arab world & 3 & 2\\
Go, my son, and fight like the honourable & 4 & 1\\
Israel is an occupying country. All settlers live in an occupied land & 5 & 3\\
Oh God, help our weak brothers and take revenge on the unjust Jews & 6 & 4 \\
\#Israel\_is\_the\_real\_terrorism \#Israel\_is\_criminal & 7 & 3\\
Jerusalem is ours and you are child killers & 8 & 3\\
God damn the Jews & 9 & 8\\
The Arab atheist is the stupidest creature on earth & 10 & 8\\
The dirtiest people in the world are the Indian people & 10 & 10\\
\bottomrule
\end{tabular}
\caption{Example of deleted posts by FB and the votes of annotators on the removal of each based on FBCS and their personal opinion (PO). \textit{*Disclaimer:} Some examples contain hate speech. This does not reflect the authors' opinions.}
\label{tab:examples}
\end{table}

\section{Discussion}

\subsection{Answering the Research Questions}

In this study, we investigated the claim of Arab activists that FB (and its algorithms) was unfairly deleting content based on alleged violations of the platform's community standards. 
Our investigation shows a clear disagreement between how FB implements its moderation practices in Arabic content (especially posts dealing with Palestine) and what Arab users consider to be harmful content.
Concerning the first three RQs, our findings indicate that (1) the majority of Arab annotators do not find that the deleted posts by FB violate the platform's community standards (RQ1); (2) the annotators also find that most of the posts should not be deleted, according to their personal opinion, even when some of the posts violate an aspect of the FBCS (RQ2); and (3) the annotators mostly agree that posts containing hate speech violate the FBCS and should be removed, but they disagree on other topics, especially those that are pro-Palestine (RQ3).

These findings open the discussion on our difficult question: Who should set and interpret the moderation guidelines of a global social media platform that serves many users of different and contradictory views? In the following, we discuss our findings and their implications in more depth, hoping to build some arguments on this question.

\subsection{Is There Evidence of Bias?}

Our findings provide evidence that, in this case, a large proportion of the posts removed by FB were not in violation of its community standards, at least according to Arab annotators. For almost 30\% of the posts that were removed, 10 out of 10 annotators assessed that they did not violate any of FB's standards (Figure~\ref{fig:removebyN}). Our findings quantitatively confirm the recent studies and reports on the same issue \cite{elmimouni2024shielding,abokhodair2024opaque,hrw_report_meta}.
The findings could be explained through the concept of cultural sensitivities \cite{hakkila_design_2020}, in that FB's ML algorithms or human moderators interpret the content of the posts differently than it was intended by the authors of the posts.
For example, one of the deleted posts in Table~\ref{tab:examples} is a prayer for Sheikh Ahmed Yassin, the founder of Hamas, to rest in peace. For the annotators, this is a nonharmful post that displays common human decency. However, since one of the subaspects of the FBCS includes \textit{``support or praise for groups or individuals involved in terrorist activity''}, it is possible that FB interprets this intent as support for terrorism, which is not how Arab annotators see it (as shown in Example 2 in Table \ref{tab:examples}).

In contrast, after the conflict between Russia and Ukraine occurred, it was leaked that FB allowed violent content against Russian invasion \citet{vengattil2022facebook}. This behaviour might highlight the double standards for dealing with content during conflicts and that social media platforms are taking certain sides. So, there is accumulating evidence of the dominant platforms being less politically neutral than they perhaps should be, or even aim to be. The sources of bias are unexplored, but they may not necessarily be the result of malicious or partial decision making; it also is possible that they are implicit and inadvertent, stemming from cultural factors and lack of understanding about marginalised communities. Therefore, it is best to avoid hasty conclusions, although users certainly have the right to \textit{express disappointment} about social media platforms' moderation standards.

\subsection{Theoretical Implications}

From a regulatory point of view, two extreme scenarios can be considered for the future of social media platforms. The first extreme is the ``open platform'', in which policies such as equal treatment of political opinions allow users to easily communicate and express their beliefs without interference from the platform (this scenario is based on the ``open Internet'' paradigm \cite{wikipedia_net_2021}). The other extreme is the ``closed platform,'' in which the platform favours certain uses, restricts access to certain information, artificially lowers the ranking of sensitive content, or explicitly filters it out.

A critical limitation of our study is that we cannot determine whether the disagreement with FB's moderation stems primarily from cultural differences between Arab and non-Arab perspectives, or represents a broader disconnect between users generally and FB's moderation practices. Without comparative data from non-Arab participants, we cannot isolate whether these findings reflect specific Arab cultural interpretations of the FBCS or point to more universal issues with FB's implementation of its standards. This distinction has important implications for platform design--if the gap is primarily cultural, it suggests a need for more culturally-sensitive moderation approaches; if it reflects a broader user-platform disconnect, it calls for fundamental changes to how moderation standards are developed and applied globally.

The issue of controlling information is not only about adhering to the law--it is also about user perceptions. Whether certain groups of users feel welcome on a given platform or not will inarguably affect their willingness to use that platform. Hence, the advocates of platform neutrality claim that political neutrality supports free speech and promotes democratic participation on a global scale. 
In actuality, the platform's design choices need to consider various viewpoints. 
On the one hand, there is a genuine need to automatically detect harmful content, such as misinformation, fake news, abuse, and hate speech. On the other hand, there has to be room for discussion on sensitive, controversial, and polarising topics--even when the discussion involves heated debate and unpopular opinions that could be falsely flagged as being against the community norms. The platform also has to ensure that human moderators it employs (or annotators of training data of moderation models) are trained to be objective and do not inject their own political beliefs into the moderation decisions. 

Political mobilisation on social networks is likely to continue to gain momentum. As evidenced by the downvoting campaign by pro-Palestinian users, people who perceive unfairness can inflict reputation damage to social media platforms via their activist campaigns. Not all of this mobilisation is carried out in good faith, but some actors use harmful tactics such as spreading rumours and misinformation that can become regarded as truth \cite{margetts_political_2017}. The main challenge remains to design effective mechanisms that are \textit{perceived} fair by all of its users. Interestingly, even when the platform's moderation design might be objective, it may be that a certain proportion of users still think that the platform is treating them unfairly. Understanding this subset of users remains an important direction for research on online communities and online user experience.

\subsection{Design Implications}

Fair treatment of users is a core tenet in the design of social media platforms, as fairness (or lack thereof) affects all social and political activity within the platform. Thus, for building sustainable online communities, fairness is a vital design consideration. Developers, designers, and content moderators working for platforms such as FB can benefit from insights into various user groups' perceptions of fair moderation when developing content moderation systems that incorporate sensibility to different cultures, languages, and political events. Because conflicts evolve rapidly and are by nature controversial, content moderation guidelines need to be more adaptive. Addressing this challenge involves both technical and human factors, such as adapting algorithmic content moderation based on near-real-time feedback from various communities.

In general, the risks of content moderation from the perspective of marginalised communities are plentiful. For example, platforms can misidentify rapid-pace tweeting during confrontations as spam, consequently blocking activist accounts and hiding tweets from public display \cite{dwoskin_facebooks_2021}. Posts about racial injustice can be classified as offensive content. Similarly, hate speech detection algorithms can associate legitimate hashtags with terrorist groups \cite{dwoskin_facebooks_2021}. These examples emphasise what can go wrong when moderating political content and how errors in this process can result in limiting activist voices and people's freedom of expression. What is needed from platforms, then, is systematic mapping of risks associated with suppressing the voice of users who have not violated the platform's rules. These efforts should lead to the design of systems that are capable, in real-time, to correct errors made by the algorithm. Most likely, more manual supervision is required, along with the design of processes that prevent \textit{real-time} bias during evolving conflicts.

As aptly formulated by Aal \cite{aal2024influence}, ``working closely with local people to understand their everyday use and appropriation of social media'' is not only a ``nice-to-have'' feature but a strict requirement in designing moderation systems that are perceived globally fair.

\subsection{Implications for HCI Researchers Studying Social Media Justice}
Because self-regulation is an opaque process based essentially on trust, it is uncertain whether self-regulation of platforms is adequate to address concerns about fairness. Although trust towards a third party with no vested interest--economic or political--would logically be established, organisations such as FB \textit{do} have vested interests. They are legal entities with the stature of publicly traded companies, meaning that their primary responsibility is towards their shareholders, not to the public.
Moreover, they are typically established in the US, hence (perhaps implicitly) driven by American values. These values may emerge in designing policies that mean well, but nonetheless include implicit and unconscious biases arising from the cultural environment \cite{hakkila_design_2020}. Hence, platforms may require more governance and regulation than previously assumed \cite{schwarz_facebook_2019}, including third-party audits.

The governments should continue carefully monitoring the design choices made by social media platforms regarding (1) political neutrality (guaranteed fair treatment of different political beliefs), (2) transparent standards and moderation policies (that are audited and validated by independent third parties with no vested interest), (3) lack of censorship based on political beliefs or other opinions falling in the domain of free speech (including validated reasoning for excluding keywords and hashtags with political nature) and (4) ensuring low barriers to entry and use for all users regardless of their characteristics such as geographic location. Ideally, platforms would incorporate these ideals in their community ideals and ensure that their moderation systems apply them consistently.

Another crucial consideration is that, at a time when conventional polling methodologies are considered to decrease their effectiveness \cite{margetts_political_2017}, ensuring accurate reflection of political diversity enables researchers to collect valid datasets that actually represent public opinions on sensitive issues. In contrast, unnecessary censorship would contaminate such datasets, which can have harmful effects on decision-making processes. Therefore, researchers should be aware of selection bias due to content deletion when performing social media analysis on sensitive topics.

\subsection{The Palestinian Context and Its Implications}

The Palestinian case presents unique challenges for content moderation that extend beyond typical concerns of hate speech or misinformation. First, the historical context of occupation and displacement means that even factual descriptions of events or expressions of solidarity can be interpreted differently by various stakeholders. Second, the power asymmetry between Palestinians and Israelis manifests itself in digital spaces through uneven access to platform governance mechanisms and different levels of algorithmic visibility.

Our findings suggest that FB's current moderation approach may inadequately account for these complexities. The high rate of disagreement between Arab annotators and FB's moderation decisions indicates a potential systematic bias in how platform policies interpret and regulate Palestinian-related content. A very recent research report conducted by the BBC shows that FB continues to suppress the reach of Palestinian posts until the time of writing this paper \cite{FB_restrict_news_2024}.
This raises broader questions about how global platforms can fairly moderate content in contexts of ongoing conflict, where the line between political expression and prohibited content may be particularly blurred.

These insights from the Palestinian context offer valuable lessons for improving content moderation globally: (1) The need for deeper understanding of historical and political contexts when developing moderation policies, (2) the importance of including marginalized voices in the development and implementation of community standards, (3) the value of transparency in moderation decisions, particularly in politically sensitive contexts, and (4) the potential role of local expertise in content moderation for specific regions or conflicts.

\subsection{Limitations and Future Research}

Regarding the survey responses, 93\% of the participants described FB as ``highly biased''. This result is likely due to our sample being skewed towards those who experience FB is treating them unfairly, as we only surveyed users who got their posts deleted by FB. Thus, generalising the results to the general population of FB users or Arab users in FB would require more research. Nonetheless, findings do indicate concerning results because the vast majority of participants whose content was deleted by FB perceive this moderation as a form of discrimination.

This study focused solely on Arab users' perceptions, which presents a limitation in interpreting our results. While we found clear disagreement between Arab users and FB's moderation decisions, we cannot determine whether non-Arab users would interpret these same posts differently or share similar concerns about FB's implementation of its standards. This limitation is fundamental to understanding the nature of the problem--is FB's moderation misaligned with users generally, or specifically with Arab cultural interpretations? A comparative study with Western users would be valuable, but poses several methodological challenges. These include: \textit{whether to use translated versions of our data} (which might lose cultural context), \textit{whether to collect a new set of deleted posts in Western languages} (complicating direct comparisons), and \textit{how to account for greater ideological diversity among Western users regarding the Palestine-Israel conflict compared to the more unified stance in Arab regions} \cite{almasri2021assessing}. Despite these challenges, such comparative work is crucial for understanding whether FB's moderation practices need cultural adaptation or more fundamental reform.

Also, since our study was related to the Palestine-Israel conflict, one of the limitations is to obtain samples of posts only from one side of this conflict. It would be interesting to obtain other samples from the Israeli side, potentially in Hebrew, and investigate user perception of the FBCS. Future work is encouraged to compare these views if it is possible to collect deleted posts on this matter from the pro-Israeli side. In addition to the Arab Spring and the Palestine-Israeli conflict, social media plays a key role in other Middle Eastern developments, such as the Syrian war, protests in Iran, and so on \cite{aal2024influence}. User studies understanding groups that play a role in these developments would be vital to the accumulation of progressive knowledge.

Finally, the sample size we collected could be seen limited (n=448). This is a limitation that is hard to overcome, because reaching FB users who got their posts banned and encouraging them to share them is not an easy task. The ideal scenario for such study is to be conducted by FB itself, where FB has the log of all banned posts and can apply an extensive quantitative and qualitative analysis to investigate our RQs above in more depth. We hope that our findings here can act as motivation for FB and other platforms to conduct such studies to improve the perception of their platforms and increase the inclusiveness of different communities. 

\section{Conclusion and Future Work}

In this study, we investigated the implementation of Facebook's community standards on Arabic posts following the Palestine-Israel conflict in May 2021. We collected 448 Arabic posts that had been moderated by FB, most of them related to the said conflict. Each post was then assessed by Arab annotators to assess if the post's content violated any of Facebook's community standards. Our findings indicate a large gap between the views of FB and Arab annotators on the posts' violation of the FBCS. We observed that pro-Palestine posts were not found to violate the FBCS by Arab annotators, but the annotators considered posts containing hate speech to be indeed violating the standards. Our findings have multiple implications, especially on the responsibility of social media companies to increase their efforts to ensure the inclusion of different user views on their platforms.

For future work, it would be highly valuable to explore how the FBCS is perceived across various communities, including Western societies on various issues, where greater variation in opinions is expected. Similarly, studying understudied communities, such as those in Asian and African countries, would provide important insights. We believe that our study offers a design framework that can be adapted to such investigations on multiple topics and communities. Finally, applying similar studies on other platforms, such as TikTok and X, would be highly important.

\bibliographystyle{ACM-Reference-Format}
\bibliography{references}

\end{document}